\documentclass[a4paper,aps,pra,showpacs,singlecolumn,notitlepage,11pt]{revtex4-1}			
   
\usepackage[utf8]{inputenc}  
\usepackage[T1]{fontenc}     
\usepackage[british]{babel}  
\usepackage{lmodern}  
\usepackage[scaled=1.03]{inconsolata} 
\usepackage[usenames,dvipsnames]{color} 
\usepackage[colorlinks,citecolor=blue,linkcolor=magenta,urlcolor=blue]{hyperref}  
\usepackage{graphicx} 
\usepackage[babel]{microtype}  
\usepackage{amsmath,amssymb,amsthm,bm,mathtools,amsfonts,mathrsfs,bbm,dsfont} 
\usepackage{xspace}  
\usepackage{multirow}
\usepackage{subcaption}

\usepackage{physics}
\newcommand{\id}{\ensuremath{\mathds{1}}}
\usepackage{bbold}

\newcommand{\bbC}{\ensuremath{\mathbb{C}}}

\newcommand{\hilb}[1]{\mathscr{#1}}

\newcommand{\ent}[1]{\mathsf{#1}}
\newcommand{\HH}{\hilb{H}}

\newcommand{\meas}[1]{\mathscr{#1}}

\newtheoremstyle{mystyle}
  {6pt}
  {6pt}
  {\normalfont}
  {0pt}
  {\bf}
  {.}
  { }
  {}

\theoremstyle{mystyle}
\newtheorem{theorem}{Theorem}

\newtheorem{corollary}{Corollary}
\newtheorem{observation}{Observation}

\newtheorem{example}{Example}

\newcommand{\dc}{\meas{C}_\textrm{DC}}
\newcommand{\gdc}{\meas{C}_\textrm{GDC}}

\begin{document}
\nonfrenchspacing
\title{Genuine Distributed Coherence}

\author{T. M. Kraft}
\affiliation{SUPA and Department of Physics, University of Strathclyde, G40NG Glasgow,
United Kingdom}
\author{M. Piani}
\affiliation{SUPA and Department of Physics, University of Strathclyde, G40NG Glasgow,
United Kingdom}


\begin{abstract}
We introduce a notion of genuine distributed coherence. Such a notion is based on the possibility of concentrating on individual systems the coherence present in a distributed system, by making use of incoherent unitary transformations. We define an entropic quantifier of genuine distributed multipartite coherence for generic mixed states, and we focus on the bipartite pure-state case. In the latter case we derive necessary and sufficient conditions for the possibility of fully localizing the coherence, hence identifying the conditions for genuine distributed bipartite coherence. We analyze in detail the quantitative problem for the case of two-qubit pure states, identifying the states with the largest amount of genuine distributed coherence. Interestingly, such states do not have maximal global coherence nor maximal coherence rank.
\end{abstract}

\maketitle

\section{Introduction}

Starting with the seminal references ~\cite{aberg_quantifying_2006,baumgratz_quantifying_2014,levi_quantitative_2014}, quantum coherence, intended as superposition of `classical' states, has been recently formalized as a resource, making use of concepts and tools of quantum information processing (see \cite{streltsov_colloquium_2017} for a recent review). The relation  of coherence with respect to quantum correlations and quantum entanglement in bi- and multi-partite systems is of particular interest, especially given that entanglement is a manifestation of coherence at the level of distributed systems, and has already enjoyed focused effort~\cite{streltsov_measuring_2015,theurer_resource_2017,regula_converting_2017,radhakrishnan_distribution_2016,bu_distribution_2017,bu_distribution_2017-1,xi_coherence_2017,yadin_quantum_2016,ma_converting_2016,streltsov_towards_2017,chitambar_relating_2016}.

In this paper we provide a general framework for the quantification of genuine distributed coherence, based on the notion of active localization of coherence on individual systems by means of incoherent unitaries that neither create nor destroy coherence; the amount of coherence that cannot be localized is then deemed to be genuinely distributed. Our approach is inspired by Ref.~\cite{kraft_characterizing_2017}, where the authors tackled the issue of pinning down genuine multilevel entanglement, considering the possibility of focusing and splitting entanglement across levels by means of the class of unitaries that are free in the resource-theoretic approach to entanglement, that is, by means of local unitaries. 

We remark that our approach is different from that of, e.g., Refs.~\cite{radhakrishnan_distribution_2016,bu_distribution_2017,bu_distribution_2017-1, xi_coherence_2017}. In the latter references the authors consider the given distribution of local and multipartite coherence; we instead consider the (reversible) manipulation of coherence under the class of unitaries that, in a resource-theoretic approach to coherence, are deemed to be ``free operations'' that preserve the coherence that is present; most importantly, while these operations maintain constant the amount of global coherence, they may allow to `focus' it on local sites. We emphasize that our notion of localization is different from the assisted distillation of coherence~\cite{chitambar_assisted_2016}. Also, our notion of ``genuine distributed coherence'' is not related to the notion of ``genuine coherence'', with the latter being the resource in a theory of coherence based on the notion of genuine incoherent operation~\cite{vicente_genuine_2017}. Finally our genuine distributed coherence is not the same as the notion of intrinsic coherence~\cite{radhakrishnan_distribution_2016}.

\section{Coherence and entanglement}

Quantum coherence, as considered in the seminal paper~\cite{baumgratz_quantifying_2014}, is a basis-dependent concept, which can be defined for a single system. Such a single system may be composite in nature, and this is the case we will consider. 
Let us denote by $\{\ket{i}\}_i$ the fixed \emph{incoherent} basis of the system; such a basis may be singled out by the physics, for example it could be the eigenbasis of the Hamiltonian of the system, or of some other specific and relevant observable. A general quantum state of the system is deemed incoherent if it is diagonal in the incoherent basis:
\[
\varrho^{\textrm{inc}} = \sum_i p_i \ket{i}\bra{i}.
\]
The set of incoherent states will be denoted by $I$.

Any state that is not incoherent is deemed coherent (equivalently, it is said to display some quantum coherence). Notice that in the case the incoherent state is pure, the corresponding vector state is simply one of the elements of the basis $\{\ket{i}\}_i$ (up to an irrelevant phase factor). A coherent vector state is a non-trivial superposition of the element of the basis; we call coherence rank the number of terms in such a superposition~\cite{theurer_resource_2017}.

Let now $\HH_{AB}=\HH_{A}\otimes\HH_{B}$ be a $d_A\times d_B$-dimensional composite Hilbert space, used to describe the state of Alice and Bob's joint system. We define the local reference basis $\qty{\ket{i}_A}_{i=0}^{d_A-1}$ for Alice and similarly $\qty{\ket{j}_B}_{j=0}^{d_B-1}$ for Bob. These are called the local incoherent bases. The joint incoherent basis is then assumed to be given by the tensor product of the local incoherent bases: $\{\ket{ij}_{AB}:=\ket{i}_A\otimes\ket{j}_B\}_{i,j}$.

A bipartite pure state
\begin{equation}
\label{eq:purestategen}
\ket{\psi}_{AB}=\sum_{ij} \psi_{ij}\ket{ij}_{AB}
\end{equation}
is incoherent if and only if exactly one of the $\psi_{ij}$ is non-zero. Otherwise, the state is coherent and the number of non-zero coefficients is, as we mentioned already in the single-system case, the coherence rank. A pure state \eqref{eq:purestategen} is \emph{unentangled} if and only if the matrix of coefficients $\Psi=[\psi_{ij}]$ has exactly one non-zero singular value, and \emph{entangled} otherwise. The number of non-zero singular values of $\Psi$ is called the Schmidt rank. Unentangled pure states have the form $\ket{\alpha}\ket{\beta}$.

A pure state \eqref{eq:purestategen} is maximally coherent if all the coefficients $\psi_{ij}$ are non-zero and equal in modulus, that is, $|\psi_{ij}| = (d_Ad_B)^{-1/2}$ for all $i,j$. Thus, a maximally coherent state has the form
\[
\ket{\psi}_{AB}=\frac{1}{\sqrt{d_Ad_B}}\sum_{ij} e^{i\varphi_{ij}}\ket{i}_A\ket{j}_{B}.
\]
A pure state \eqref{eq:purestategen} is maximally entangled if and only if the matrix $[\psi_{ij}]$ has $\min\{d_A,d_B\}$ non-zero and equal singular values, which are then necessarily equal to $(\min\{d_A,d_B\})^{-1/2}$. Thus, a maximally entangled state has the form
\[
\ket{\psi}_{AB}=\frac{1}{\sqrt{\min\{d_A,d_B\}}}\sum_{i}\ket{\alpha_i}_A\ket{\beta_i}_{B},
\]
where $\{\ket{\alpha_i}\}$ and $\{\ket{\beta_j}\}$ are local orthonormal bases.
Notice that one can absorb any phases in the definition of the local bases. 

Both for coherence and entanglement, the concepts are generalized to the mixed-state case by simply considering convex combinations. The incoherent (unentangled) mixed states are all and only those that can be obtained by taking convex combinations of pure incoherent (entangled) states. States of the form
\begin{equation}
\label{Eq:incoherentmixedstate}
\varrho=\sum_{ij} p_{ij} \ketbra{i}\otimes\ketbra{j}
\end{equation}
that are diagonal in the joint incoherent basis constitute the set of bipartite incoherent states $I$. Unentangled states are those that can be written as convex combinations of unentangled pure states, that is, states of the form
\[
\rho_{AB}=\sum_{i}p_i\ketbra{\alpha_i}_A\otimes\ketbra{\beta_i}_{B}.
\]

Both coherence and entanglement can be considered resources in frameworks where there are certain specific limitations. We will focus in particular on unitary transformations that leave coherence invariant. For a single system, incoherent unitary operations are of the form
\begin{equation}
\label{Eq:IncoherentUnitarySingle}
U=\sum_{j} e^{i\varphi_{j}} \ketbra{\pi(j)}{j},
\end{equation}
i.e. they can be written as a phase gate and a permutation $\pi$ of the incoherent basis. These are the most general unitary operations under which the set $I$ is closed.
In the bipartite setting, incoherent unitary operations are of the form
\begin{equation}
\label{Eq:IncoherentUnitary}
U=\sum_{ij} e^{i\varphi_{ij}} \ketbra{\pi(ij)}{ij}.
\end{equation}

\section{Relative entropy of coherence}

As a quantifier of coherence we will use the \textit{relative entropy of coherence} \cite{baumgratz_quantifying_2014} that is defined by
\begin{equation}
\label{Eq:RelEntropyCoherence}
\meas{C}(\varrho):=\min_{\sigma\in I}\ent{S}(\varrho\parallel\sigma).
\end{equation}
Here,  $\ent{S}(\varrho\parallel\sigma):=\tr[\varrho\log(\varrho)]-\tr[\varrho\log(\sigma)]$ is the relative entropy, and the minimization is over all incoherent states $\sigma$. A very useful property of the relative entropy of coherence is that it is additive on tensor products,
\begin{equation}
\label{eq:Cadditivity}
\meas{C}(\varrho_A\otimes\varrho_B)=\meas{C}(\varrho_A)+\meas{C}(\varrho_B).
\end{equation}
In addition, an analytic solution to the minimization problem is known \cite{baumgratz_quantifying_2014}. The relative entropy of coherence can be expressed as
\begin{eqnarray}
\label{Eq:AnalyticExpression}
\meas{C}(\varrho)&=&\ent{S}(\varrho^{d})-\ent{S}(\varrho)
\end{eqnarray}
where $\ent{S}(\varrho)=-\tr[\varrho\log(\varrho)]$ is the von Neumann entropy and $\varrho^{d}=\sum_i \bra{i}\varrho\ket{i} \ketbra{i}$ is the totally decohered version of the state $\varrho$. It is immediate to see that the relative entropy of coherence is invariant under the action of incoherent unitary transformations, which makes it a good coherence quantifier~\cite{baumgratz_quantifying_2014}.

\section{A first look at distributed coherence}

In multipartite systems one can distinguish between different manifestations of coherence, going beyond simply detecting and quantifying coherence in the joint incoherent basis $\{\ket{ij}\}$ (see also~\cite{radhakrishnan_distribution_2016,bu_distribution_2017,bu_distribution_2017-1,xi_coherence_2017,yadin_quantum_2016,ma_converting_2016,streltsov_towards_2017}). What we are mostly interested in in this work is the relation between global coherence---that is, the coherence of the global state---and the local coherence---the coherence exhibited by the local reduced states.
\begin{example}
In the simplest case the systems are uncorrelated and their state does not contain any coherence at all, such in the case of
\begin{equation}
\ket{0}\ket{0}.
\end{equation}
\end{example}
\begin{example}
Then, there exist coherent, yet uncorrelated states. Consider the state
\begin{equation}
\ket{+}\ket{+}=\frac{1}{2}\sum_{i,j=1}^1\ket{ij},
\end{equation}
with $\ket{\pm}=(\ket{0}\pm\ket{1})/\sqrt{2}$. Here, not only the global state is coherent, but also its marginals are. In fact, the amount of local coherence is equal to the amount of global coherence, in the sense that $\meas{C}(\varrho_{AB})=\meas{C}(\varrho_{A})+\meas{C}(\varrho_{B})$.
\end{example}
A more interesting class of states are those that are globally coherent, but, due to the fact that they are entangled, have incoherent marginals. Nevertheless, in some of these cases the coherence can be concentrated on the subsystems by applying incoherent unitary operations such that the global coherence is preserved, but converted to local coherence.

\begin{example}
Consider the maximally entangled state in dimension $d\times d$
\begin{equation}
\ket{\psi_d}=\frac{1}{\sqrt{d}}\sum_{i}\ket{ii}.
\end{equation}
This state has coherence rank $d$ and its coherence is a property of the bipartite system since both marginals are maximally mixed and thus incoherent. Interestingly, all the coherence can be concentrated on one of the subsystems, say Alice, by applying an incoherent unitary operation. Indeed,
\begin{equation}
\ket{\psi_d}=\ent{CNOT}\qty[\frac{1}{\sqrt{d}}\sum_i\ket{i}\otimes \ket{0}],
\end{equation}
where $\ent{CNOT}$ is the generalized controlled-not gate (more precisely, a controlled shift) acting as $\ent{CNOT}\ket{i}\ket{j}=\ket{i}\ket{j\oplus i}$, where the addition $\oplus$ is modulo $d_B$. Notice that the coherence in the state inside the square brackets is located in Alice's system.
\end{example}

\section{A quantifier of genuine distributed coherence}
\label{sec:gmc}

Our objective is to study coherence in multipartite systems by considering entropic quantifiers to measure to what extent coherence is spread across the subsystems and to what extent it can be concentrated on the individual systems by means of incoherent unitary operations. While we will focus on the bipartite case for the sake of clarity and conciseness, essentially all of the basic definitions extend naturally to the multipartite case, and for this reason we may use the adjective ``multipartite'' even if focusing on the bipartite case.

\subsection{Distributed coherence}

In a first step we want to quantify to what extent the coherence of a state is a property of the bipartite state and not only of its marginals. We propose the following quantifier of distributed coherence (DC) to characterize the multipartite (as opposite to localized) coherence of a state:
\begin{equation}
\label{eq:CDCdef}
\meas{C}_{DC}(\varrho_{AB}):=\meas{C}(\varrho_{AB})-\meas{C}(\varrho_A\otimes\varrho_B)=\meas{C}(\varrho_{AB})-\qty[\meas{C}(\varrho_A)+\meas{C}(\varrho_B)].
\end{equation}
Again, we point out that this could as easily be defined directly / generalized for actual multipartite systems in a straightforward way; explicitly:
\begin{equation}
\begin{split}
\meas{C}_{DC}(\varrho_{A_1A_2\ldots A_n})=\meas{C}(\varrho_{A_1A_2\ldots A_n})-\meas{C}(\bigotimes\varrho_{A_i}) =\meas{C}(\varrho_{A_1A_2\ldots A_n})-\sum_i \meas{C}(\varrho_{A_i}).
\end{split}
\end{equation}
We remark that the fact that our distributed-coherence quantifier is equal both to the gap between global coherence and the sum of the local coherences, and to the gap between global coherence and the coherence of the product of the marginals, is a consequence of the additivity of the relative entropy of coherence on tensor products, Eq.~\eqref{eq:Cadditivity}. While this is beyond the scope of the present work, one can define and study gaps between the global and local coherences that are based on other coherence quantifiers, and in such a case, one would in general deal with two distinct gaps.

Recall that there is an analytic expression that can be used to express our quantifier \eqref{eq:CDCdef}
 in terms entropies of the original state $\varrho_{AB}$, its decohered version $\varrho_{AB}^{d}$ and their marginals. Inserting the expression from Eq. \eqref{Eq:AnalyticExpression}, one obtains
\begin{equation}
\label{Eq:MCquant}
\begin{split}
\meas{C}_{DC}(\varrho_{AB})&=\ent{S}(\varrho_{AB}^{d})-\ent{S}(\varrho_{AB})-\qty[\ent{S}(\varrho_A^{d})-\ent{S}(\varrho_A)+\ent{S}(\varrho_B^{d})-\ent{S}(\varrho_B)]\\
&=I_{\varrho}(A:B)-I_{\varrho^d}(A:B)\\
&\coloneqq\Delta I_{\varrho}(A:B).
\end{split}
\end{equation}
That this difference is not negative comes from the data-processing inequality~(see the comprehensive~\cite{beaudry_intuitive_2011}) related to strong-subadditivity of the von Neumann entropy, which ensures that mutual information $I_\varrho(A:B)=S(\varrho_A)+S(\varrho_B)-S(\varrho_{AB})=S(\varrho_{AB}\|\varrho_A\otimes\varrho_B)$ decreases under local operations, in particular under local projective measurements.
We recognize $\Delta I_{\varrho}(A:B)$ as a basis-dependent version of a discord quantifier based on the notion of local projective measurements, meant to capture the quantumness of correlations~\cite{ollivier_quantum_2001,modi_classical-quantum_2012}. One obtains a basis-\emph{in}dependent discord quantifier---actually, what we normally refer to as discord quantifier---by minimizing the gap $\Delta I_{\varrho}(A:B)$ over the choice of local bases~\cite{girolami_faithful_2011,modi_classical-quantum_2012}, equivalently, by optimizing over local unitaries~\footnote{We remark that, in the case where one is interested in the quantumness of correlations, a further option is that of optimizing over general local measurements (also known as POVMs) meant to ``extract'' the largest possible amount of classical correlation~\cite{piani_no-local-broadcasting_2008}.}. As we will see later, in this paper we go down another route, optimizing over arbitrary (that is, also global) incoherent unitaries. We point out that the above identity $\dc(\rho) = \Delta I_\varrho(A:B)$, together, with the data processing inequality provides an immediate and simple proof of some of the results of \cite{bu_distribution_2017} relating coherence and discord.

It is natural to consider the problem of when $\Delta I_{\varrho}(A:B)$ vanishes. One solution is certainly the trivial case when the state $\varrho_{AB}$ itself is already incoherent. Another trivial solution is the case where the state $\varrho_{AB}$ is uncorrelated, i.e., $\varrho_{AB}=\varrho_A\otimes\varrho_B$, so that the mutual information vanishes to begin with. The authors of \cite{yadin_quantum_2016}, building on techniques from~\cite{piani_no-local-broadcasting_2008}, and amending a statement made in~\cite{ollivier_quantum_2001}, derived the structure of the states $\varrho_{AB}$ for which the condition
\begin{equation}
\label{eq:onesidebasisdepdiscord}
I_{\varrho}(A:B)-I_{\varrho^{d_A}}(A:B) = 0
\end{equation}
holds, where $\varrho^{d_A}=\sum_i\ketbra{i}_A\otimes\id_B \varrho \ketbra{i}_A\otimes\id_B$. Such a structure can be understood in terms of a partitioning of elements of the incoherent basis $\{\ket{i}\}$ on $A$ into disjoint subsets, and of orthogonal projectors $\{P_a\}$ each projecting on the subspace spanned by one subset in such partitioning. Equivalently, we can speak of an orthogonal projective decomposition of the identity, where each projector is diagonal in the incoherent basis. Then the condition for \eqref{eq:onesidebasisdepdiscord} to hold is that
\[
\sum_a P^A_a \otimes \id_B \varrho_{AB} P^A_a \otimes \id_B = \varrho_{AB}
\]
and that
\[
P^A_a \otimes \id_B \varrho_{AB} P^A_a \otimes \id_B
\]
is product for every projector $P_a$.
It is immediate to realize that, in the case where one considers projective measurements on both parties, one has $\Delta I_{\varrho}(A:B)=0$ if and only if 
\begin{equation}
\label{eq:invariantstate}
\sum_a P^A_a \otimes P^B_b \varrho_{AB} P^A_a \otimes P^B_b = \varrho_{AB},
\end{equation}
and it holds both that
\[
P^A_a \otimes \id_B \varrho_{AB} P^A_a \otimes \id_B
\]
is uncorrelated for all $a$, and that
\[
\id_A \otimes P^B_b \varrho_{AB} \id_A \otimes P^B_b
\]
is uncorrelated for all $b$, for some local orthogonal projective measurements $\{P^A_a\}$  and $\{P^B_b\}$ which are diagonal in the respective local incoherent bases. Notice that such conditions imply that 
\[
P^A_a \otimes P^B_b \varrho_{AB} P^A_a \otimes P^B_b
\]
is uncorrelated for all $a$ and $b$.
We point out how this characterization covers both trivial cases mentioned above, i.e., incoherent states and product states.

In the following we will focus on pure states. It is clear that in the case of a pure state $\varrho_{AB}=\ketbra{\psi}_{AB}$ the conditions above can only be satisfied by a product state $\ket{\psi}_{AB}=\ket{\alpha}_A\ket{\beta}_B$. This is because Eq. \eqref{eq:invariantstate} implies that $P^A_a \otimes P^B_b \ket{\psi}_{AB}$ must be proportional to $\ket{\psi}_{AB}$, besides also being uncorrelated.

\subsection{Genuinely distributed coherence}

We introduce the concept of \emph{genuine distributed coherence} by taking into consideration that, in the framework of incoherent operations introduced by Baumgratz et al.~\cite{baumgratz_quantifying_2014}, the coherence present in a distributed system is invariant under incoherent unitaries~\cite{streltsov_measuring_2015}, which are considered as ``free operations'', even in the case where they are non-local. That is, (global) incoherent unitaries play the same role in coherence theory as local unitaries play in entanglement theory, at least, as mentioned, in the framework of Ref.~\cite{baumgratz_quantifying_2014}.

Taking this idea seriously, as done previously in, for example, Refs.~\cite{streltsov_measuring_2015,winter_operational_2016}, in this paper we focus on the amount of multipartite coherence that remains after a minimization of $\meas{C}_{DC}$ over all incoherent unitaries defined by Eq. \eqref{Eq:IncoherentUnitary}. This leads to the following definition of genuinely distributed coherence (with stratighforward generalization to the multipartite case):
\begin{equation}
\label{Eq:GMC}
\meas{C}_{GDC}(\varrho_{AB})=\min_{U_I} \qty[ \meas{C}(\xi_{AB})-\meas{C}(\xi_A\otimes\xi_B)]_{\xi=U_I\varrho_{AB}U_I^{\dagger}}=\min_{U_I} \Delta I_{\xi}(A:B)\big\rvert_{\xi=U_I\varrho_{AB}U_I^{\dagger}}.
\end{equation}

\section{Distributed and genuine distributed coherence for pure bipartite states}

After having defined our concepts and quantifiers in a general way---that is, for mixed multipartite states---in the previous sections, in this section we focus on pure bipartite states.

\subsection{Pure bipartite states with vanishing genuine multipartite coherence}

We have argued that the only pure bipartite state with vanishing multipartite coherence $\meas{C}_{DC}$ are factorized states. This implies that the only pure bipartie states $\ket{\psi}_{AB}$ with vanishing \emph{genuine} distributed coherence $\meas{C}_{GDC}$ are those that can be decorrelated by means of an incoherent unitary. We now derive necessary and sufficient conditions for this to be possible.

Given a pure state $\ket{\psi}$, we can expand it in the incoherent basis,
\begin{equation}
\ket{\psi}=\sum_{ij}\psi_{ij}\ket{ij}=\sum_{ij}\abs{\psi_{ij}}e^{i\varphi_{ij}}\ket{ij},
\end{equation}
where $\bbC\ni\psi_{ij}=\abs{\psi_{ij}}e^{i\varphi_{ij}}$. Then, a state $\ket{\psi}$ can be decorrelated by incoherent unitaries $U_I$ if and only if
\begin{equation}
\max_{U_I,\ket{ab}}\abs{\bra{ab}U_I\ket{\psi}}=1 , 
\end{equation}
where $\ket{ab}=\sum_{ij}a_ib_j\ket{ij}=\sum_{ij}\abs{a_i}\abs{b_j}e^{i(\alpha_i+\beta_j)}\ket{ij}$. Recall, that incoherent unitaries can be written as a combination of a phase gate and a permutation in the incoherent basis (see Eq.~\eqref{Eq:IncoherentUnitary}). Thus, one has
\[
\ket{\psi'} = U_I\ket{\psi} = \sum_{ij}\abs{\psi_{\pi(ij)}}e^{i\varphi'_{ij}}\ket{ij}.
\]
We remark that, thanks to the freedom in the phases of the incoherent unitary, the phases $\varphi'_{ij}$ can be chosen arbitrarly, when optimizing over $U_I$.
One therefore has,
\[
\begin{split}
\max_{U_I}\abs{\bra{ab}U_I\ket{\psi}}
&=\max_{\pi,{\varphi_{ij}'}}\abs{\sum_{ij}\abs{\psi_{\pi(ij)}}\abs{a_i}\abs{b_j}e^{i(-\alpha_i-\beta_j+\varphi'_{ij})}}\\
&\leq \max_{\pi}\sum_{ij}\abs{\psi_{\pi(ij)}}\abs{a_i}\abs{b_j}
\end{split}
\]
where the last inequality---coming from the triangle inequality---can be saturated by a suitable choice of phases $\varphi_{ij}'$, specifically $\varphi_{ij}'=\alpha_i+\beta_j$.

Then we are left with optimizing
\begin{equation}
\max_{\pi,\ket{ab}}\sum_{ij}\abs{\psi_{\pi(ij)}}\abs{a_i}\abs{b_j}
=\max_{\pi}\qty[ \|\Psi^\textrm{abs}_\pi\|_\infty],
\end{equation}
where $\Psi^\textrm{abs}_\pi=\left[\abs{\psi_{\pi(ij)}}\right]$ is the matrix of the moduli of the coefficients $\psi_{ij}$, rearranged according to the permutation $\pi$, and $\|\cdot\|_\infty$ indicates the largest singular value. From this the following observation follows. 
\begin{theorem}
A bipartite pure state $\ket{\psi}$ of dimension $d_A\times d_B$ with coefficients $\psi_{ij}\in\bbC$ has $\meas{C}_{GDC}(\ketbra{\psi})=0$ if and only if $\max_{\pi}\qty[\|\Psi^\textrm{abs}_\pi\|_\infty ]=1$, where $\Psi^\textrm{abs}_\pi=\left[\abs{\psi_{\pi(ij)}}\right]$, and the maximization is over all permutations of the pairs $(i,j)$. An equivalent condition is that there is a permutation $\pi$ such that $\Psi^\textrm{abs}_\pi$ has rank equal to one.
\end{theorem}

\begin{observation}
\label{obs:maxentexact}
Any two-qubit maximally entangled state $\psi$ has vanishing genuine distributed coherence. This is clear, once one considers that the matrix of coefficients $\Psi=[\psi_{ij}]$ is in such a case proportional to a unitary matrix, whose rows and columns are orthogonal vectors, so that necessarily $\psi^*_{00}\psi_{01}=-\psi^*_{10}\psi_{11}$, and hence $\abs{\psi_{00}}\abs{\psi_{01}}-\abs{\psi_{10}}\abs{\psi_{11}}=0$; this proves that there is a permutation $\pi$ such that $\Psi^\textrm{abs}_\pi$ has rank equal to one.
\end{observation}

We see that for two qubits, maximal entanglement is not compatible with the presence of genuine distributed coherence. Is this the case for all maximally entangled states in any local dimension? The following proves that it is not.

\begin{corollary}
Any pure state $\ket{\psi}$ such that $\Psi$ has a number of non-vanishing entries equal to a prime number striclty larger than $\max\{d_A,d_B\}$ has non-zero genuine distributed coherence.
\end{corollary}
This is because, for $\Psi^\textrm{abs}_\pi$ to have rank one, that is, to be of the form $\ket{a}\bra{b}$, it must be that the number of its non-zero entries is either less or equal to  $\max\{d_A,d_B\}$, or not a prime number.

\begin{example}
The two-qutrit maximally entangled state
\[
\frac{1}{\sqrt{3}}\left(\ket{+}\ket{+}+i\ket{-}\ket{-}+\ket{2}\ket{2}\right)
\]
with $\ket{\pm}=(\ket{0}\pm\ket{1})/\sqrt{2}$, has genuinely distributed coherence, since it has five non-vanishing coefficient when expressed in the standard othonormal basis $\{\ket{i}\ket{j}\}$.
\end{example}

Despite the fact that computing the maximal singular value of the matrix $\Psi^\textrm{abs}_\pi$ is rather easy, there still remains the problem of optimizing over the permutations of the indices. An upper bound on the number of arrangements of the coefficients that could potentially lead to different singular values is given by
\begin{equation}
\label{Eq:Num1}
N=\frac{(d_A\times d_B)!}{\prod_{i,j} (i+j-1)}.
\end{equation}
If one is only interested in whether or not there is an arrangement, such that $\text{rank}(\Psi_\pi)=1$, the number of arrangements that one has to test is at most
\begin{equation}
\label{Eq:Num2}
N'=\frac{(d_A + d_B -2 )!}{(d_A-1)! \times (d_B-1)!}.
\end{equation}
As mentioned, our approach is insipired by the problem of characterizing high-dimensional entanglement tackled in \cite{kraft_characterizing_2017}; a detailed proof and discussion of Eqs. \eqref{Eq:Num1} and \eqref{Eq:Num2} can be found therein.

For the case of two qubits the optimization over the permutations of coefficients can, however, easily be performed. Observe that if $\Psi^\textrm{abs}_\pi$ has rank one, it can be written as $\ketbra{a}{b}$ and we can assume without loss of generality that $a_0>a_1$ and $b_0>b_1$ (due to the freedom of absorbing the local permutations $\psi_{0j}\leftrightarrow\psi_{1j}$ and $\psi_{i0}\leftrightarrow\psi_{i1}$, which cannot change $\dc$, since they preserve both the global and the local coherences). Hence, it is optimal to permute the largest element in the upper left entry and the smallest in the lower right entry. The position of the intermediate values does not matter, since the rank is invariant under transposition. Hence we arrive at the following observation.
\begin{corollary}
A generic two-qubit pure state $\ket{\psi}=\psi_{00}\ket{00}+\psi_{01}\ket{01}+\psi_{10}\ket{10}+\psi_{11}\ket{11}$ has zero genuine multipartite coherence in the standard computational basis corresponding to this expansion if and only if
\begin{equation}
\det\mqty[\abs{\psi_{\max}} & \abs{\psi_{1}} \\ \abs{\psi_{2}} & \abs{\psi_{\min}}]=\abs{\psi_{\max}}\abs{\psi_{\min}}-\abs{\psi_{1}}\abs{\psi_{2}}=0,
\end{equation}
where $\psi_{\max}$ is the largest coefficient, $\psi_{\min}$ the smallest, and $\psi_{1,2}$ are the remaining two coefficients.
\end{corollary}

\subsection{Distributed coherence of pure two-qubit states}

In this subsection we illustrate the concept of distributed coherence of Section \ref{sec:gmc} by evaluating its quantifier $\meas{C}_\textrm{DC}$ for generic two-qubit pure states. Again, we consider the generic form of a pure state,
\begin{equation}
\ket{\psi}=\sum_{ij}\psi_{ij}\ket{ij}.
\end{equation}
As noticed before, $\meas{C}_\textrm{DC}$ coincides with the difference in the mutual information, given by
\begin{equation}
\Delta I(A:B)=\ent{S}(\varrho_A)+\ent{S}(\varrho_B)-\ent{S}(\varrho_{AB})-\ent{S}(\varrho^d_A)-\ent{S}(\varrho^d_B)+\ent{S}(\varrho^d_{AB}).
\end{equation}
First, note that $\ent{S}(\varrho_{AB})=0$, since the global state is pure. For the other entropies one obtains
\begin{eqnarray}
\ent{S}(\varrho_A)+\ent{S}(\varrho_B)&=&2\Bigg[-\qty(\frac{1}{2}+\sqrt{\frac{1}{4}-\abs{\det(\Psi)}^2})\log_2\qty(\frac{1}{2}+\sqrt{\frac{1}{4}-\abs{\det(\Psi)}^2})\notag\\
&&\quad-\qty(\frac{1}{2}-\sqrt{\frac{1}{4}-\abs{\det(\Psi)}^2})\log_2\qty(\frac{1}{2}-\sqrt{\frac{1}{4}-\abs{\det(\Psi)}^2})\Bigg]\notag\\
&=&2h\bigg(\frac{1}{2}+\sqrt{\frac{1}{4}-\abs{\det(\Psi)}^2}\bigg)\\
\ent{S}(\varrho^d_A)+\ent{S}(\varrho^d_B)&=&-(\abs{\psi_{00}}^2+\abs{\psi_{01}}^2)\log_2(\abs{\psi_{00}}^2+\abs{\psi_{01}}^2)-(\abs{\psi_{10}}^2+\abs{\psi_{11}}^2)\log_2(\abs{\psi_{10}}^2+\abs{\psi_{11}}^2)\notag\\
&&-(\abs{\psi_{00}}^2+\abs{\psi_{10}}^2)\log_2(\abs{\psi_{00}}^2+\abs{\psi_{10}}^2)-(\abs{\psi_{01}}^2+\abs{\psi_{11}}^2)\log_2(\abs{\psi_{01}}^2+\abs{\psi_{11}}^2)\notag\\
\ent{S}(\varrho_{AB}^d)&=&-\sum_{ij}\abs{\psi_{ij}}^2\log_2(\abs{\psi_{ij}}^2),
\label{eq:C_DC}
\end{eqnarray}
where, we recall,
\begin{equation}
\Psi= \mqty[\psi_{00}&\psi_{01}\\\psi_{10}&\psi_{11}]
\end{equation}
is the matrix of coefficients in the standard computational/incoherent basis, and $h(p):=-p\log p -(1-p)\log_2(1-p)$ is the binary entropy.

The above constitutes a generic expression of $\meas{C}_\textrm{DC}$ for any two-qubit pure state. It simplifies substantially for, e.g., a maximally entangled state. In the latter case, as mentioned already in Observation~\ref{obs:maxentexact}, the matrix of coefficients $\Psi = [\psi_{ij}]$ is proportional to a unitary, more precisely $\Psi=U/\sqrt{2}$, so that $|\det(\Psi)|=1/2$, and the reduced states are maximally mixed. Thus $\Delta I(A:B)=\ent{S}(\varrho_{AB}^d)$. Since  $\Psi=U/\sqrt{2}$, with the columns and rows of $U$ orthonormal, we have that
\[
\begin{split}
-\sum_{ij}\abs{\psi_{ij}}^2\log_2(\abs{\psi_{ij}}^2)
&= -2 (|\psi_{00}|^2 \log_2 |\psi_{00}|^2 + |\psi_{01}|^2 \log_2 |\psi_{01}|^2 ) \\
&= -2 \left(\frac{p}{2} \log_2 \frac{p}{2} + \frac{1-p}{2} \log_2 \frac{1-p}{2} \right)\\
&= 1 + \left(- p \log_2 p - (1-p)\log_2 (1-p)\right)\\
&= 1 + h(p),
\end{split}
\]
with $p=2|\psi_{00}|^2=2|\psi_{11}|^2$.

Thus the maximal amount of coherence for a maximally entangled state can simply be computed. It turns out that the state
\begin{equation}
\ket{\psi}=\frac{1}{\sqrt{2}}(\ket{0}\ket{+}+\ket{1}\ket{-})=\ket{\bullet\hspace*{-1mm}-\hspace*{-1mm}\bullet},
\end{equation}
which is a two-qubit graph state \cite{hein_entanglement_2006}, has the maximum amount of distributed coherence, $\Delta I(A:B)=2$. Nonetheless, as we have seen in Observation~\ref{obs:maxentexact}, a maximally entangled two-qubit state has zero genuine distributed coherence because it can be decorrelated by an incoherent unitary; indeed, in this specific case, by applying a controlled phase gate the state $\ket{+}\ket{+}$ is obtained.

\subsection{Genuine distributed coherence of pure two-qubit states}

We now tackle the calculation of $\meas{C}_\textrm{GDC}$ of a given two-qubit pure state, obtained by minimizing $\meas{C}_\textrm{DC}$ over incoherent unitaries. Since the coherence $\meas{C}(\rho_{AB})$ is invariant under the action of the incoherent unitary operation, we are left with maximizing the sum of the local coherences, that is, with calculating
\begin{equation}
\max[\meas{C}(\varrho_A)+\meas{C}(\varrho_B)]=\max\qty{\ent{S}(\varrho^d_A)+\ent{S}(\varrho^d_B)-[\ent{S}(\varrho_A)+\ent{S}(\varrho_B)]}.
\label{eq:maxlocalcoherence}
\end{equation}
First, note that only the maximization of the local coherences of $\varrho_A$ and $\varrho_B$ depends on the phases of the coefficients $\psi_{ij}$. The maximization of these terms is equivalent to the minimization of the square of the absolute value of the determinant of $\Psi$,
\[
\abs{\det(\Psi)}^2=|\psi_{00}\psi_{11}-\psi_{01}\psi_{10}|^2=\abs{\psi_{00}}^2\abs{\psi_{11}}^2+\abs{\psi_{01}}^2\abs{\psi_{10}}^2-2\Re\qty{\psi_{00}\psi_{01}^*\psi_{10}^*\psi_{11}}.
\]
Here, the minimum is obtained if all the phases in the last term cancel, i.e. if the product is rotated to the positive real axis. Therefore it is justified to assume that all the $\psi_{ij}$ are real and positive, that is, to work with $\Psi^\textrm{abs}$ rather than $\Psi$; indeed, we can achieve this by means of the phase freedom in the incoherent unitary.

Having optimized over the phases of the incoherent unitary, we now need to consider the optimization over permutations $\pi$ of the pairs $(i,j)$. Given the expressions in Eqs.~\eqref{eq:C_DC}, it is immediate to realize that it is sufficient to consider only the permutations given by the identity, by $(0,1)\leftrightarrow (1,1)$, and by $(1,0)\leftrightarrow (1,1)$. That is, the 
three arrangements of coefficients that could potentially lead to different values of the quantifier are the following:
\begin{eqnarray}
\Psi= \mqty[\psi_{00}&\psi_{01}\\\psi_{10}&\psi_{11}],\quad\Psi'= \mqty[\psi_{00}&\psi_{11}\\\psi_{10}&\psi_{01}],\quad\Psi''= \mqty[\psi_{00}&\psi_{01}\\\psi_{11}&\psi_{10}].
\label{eq:rearrangements} 
\end{eqnarray}

Thus, we have found that, for any given two-qubit pure state, one can compute the value of $\meas{C}_{\textrm{GDC}}$ by evaluating the quantities in~\eqref{eq:C_DC} with the use of the absolute values of the amplitudes, and for all the rearrangements~\eqref{eq:rearrangements}, then picking the arrangement that realizes \eqref{eq:maxlocalcoherence}.

By optimizing numerically~\footnote{We used the function NMaximize in Wolfram Mathematica 11.0.1.0.} over the amplitudes $\psi_{ij}$ for a two-qubit pure state, we observe that the largest amount of genuine distributed coherence is achieved by pure states with coherence rank equal to three, rather than maximal (that is, four). More precisely, we find that a state with the largest genuine distributed coherence is
\begin{equation}
\label{eq:maximalGDC}
\ket{\psi}=\frac{1}{\sqrt{3}}(\ket{00}+\ket{01}+\ket{10})=\frac{1}{\sqrt{3}}(\sqrt{2}\ket{0}\ket{+}+\ket{1}\ket{0}),
\end{equation}
which has reduced states
\[
\rho_A=\rho_B = \frac{2}{3}\ketbra{+}+\frac{1}{3}\ketbra{1}.
\]
Such a state has global coherence $\meas{C}(\ketbra{\psi})=\log_2 3$ and local coherences $\meas{C}(\rho_A)=\meas{C}(\rho_B)=h(1/3)-h((3+\sqrt{5})/6)$, so that it has distributed coherence
\[
\log_2 3 - 2(h(1/3)-h((3+\sqrt{5})/6))\approx 0.8485,
\]
which can not be further decreased by incoherent unitaries, as evident from Eqs. \eqref{eq:C_DC} and from the discussion in this subsection. States that have the same amplitudes as $\ket{\psi}$, up to phases and to relabelling of the elements of the incoherent basis, have the same genuine distributed coherence and even the same distributed coherence.

\begin{figure}[t!]
    \centering
    \begin{subfigure}[t]{0.45\textwidth}
        \centering
        \includegraphics[height=2.2in]{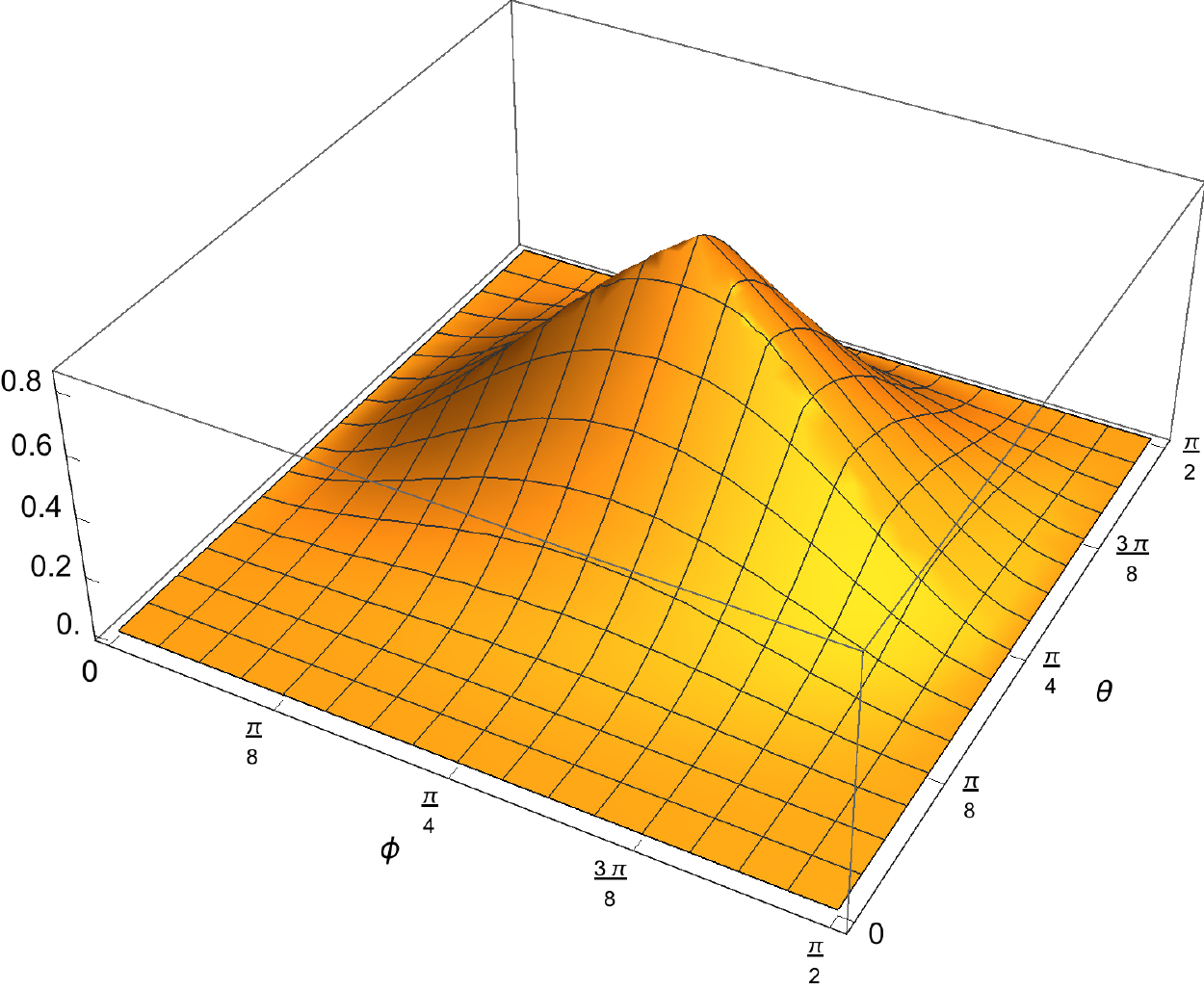}
    \end{subfigure}
    ~
    \begin{subfigure}[t]{0.45\textwidth}
        \centering
        \includegraphics[height=2.2in]{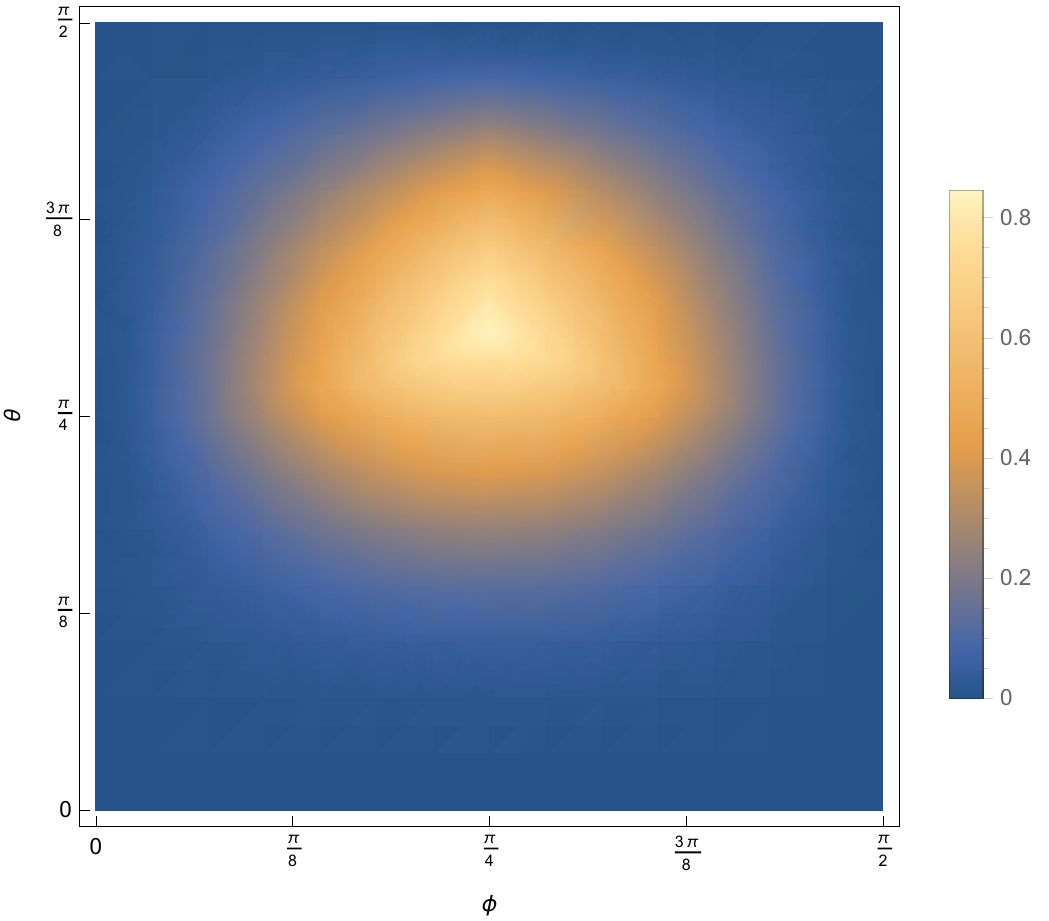}
    \end{subfigure}
    \caption{Genuine distributed coherence $\meas{C}_{\textrm{GDC}}$ for the class of states described in Eq.~\eqref{eq:rankthreeGDC}.}
    \label{fig:maxGDC}
\end{figure}

More in general, taking into account our discussion on the optimization of phases, so that only real positive $\psi_{ij}$ need to be considered to find a maximum for $\meas{C}_{\textrm{GDC}}$, one is led to consider the class of rank-three states characterized by points in the first octant on the three-dimensional unit sphere, which can be written using spherical coordinates:
\begin{equation}
\label{eq:rankthreeGDC}
\ket{\psi}=\sin(\theta)\cos(\phi)\ket{00}+\sin(\theta)\sin(\phi)\ket{01}+\cos(\theta)\ket{10},
\end{equation}
where $\theta,\phi\in\qty[0,\pi/2]$.
In Figure \ref{fig:maxGDC} we have plotted the genuine distributed coherence $\meas{C}_{\textrm{GDC}}$ as a function of $\theta$ and $\phi$, which shows graphically how, within this class, the state \eqref{eq:maximalGDC} is optimal. 

It is worth remarking that the state Eq. \eqref{eq:maximalGDC} that has the largest amount of genuinely distributed coherence has a structure similar to that of the four-qubit state that was shown to have the largest amount of genuine multilevel entanglement (see Ref. \cite{kraft_characterizing_2017}, in particular Observation 2):
\begin{equation}
    \ket{\xi}=\frac{1}{\sqrt{3}}\qty(\ket{00}+\ket{11}+\ket{22})_{AB}.
\end{equation}
This state cannot be reproduced by two pairs of (potentially entangled) qubits together with arbitrary local unitary operations on Alice's and Bob's qubits respectively; that is,
\[
\ket{\xi}_{AB}\neq U_{A_1A_2}\otimes V_{B_1B_2}\ket{\psi_1}_{A_1B_1}\ket{\psi_2}_{A_2B_2},
\]
for any two qubit states $\ket{\psi_1}$ and $\ket{\psi_2}$, and any two-qubit unitaries $U$ and $V$.
Interestingly, however, the state $\ket{\xi}$ can be produced from the state \eqref{eq:maximalGDC}. Think of the latter as being the state of the two qubits held by Alice, and let such qubits each interact independently with one qubit of Bob, initially prepared in the state $\ket{0}$, via a $\ent{CNOT}$, so to obtain
\begin{align*}
    &\quad(
    \ent{CNOT}_{A_1B_1}\otimes \ent{CNOT}_{A_2B_2})\bigg[\frac{1}{\sqrt{3}}\qty(\ket{00}+\ket{01}+\ket{10})_{A_1A_2}\otimes\ket{00}_{B_1B_2}\bigg]\\
&=\frac{1}{\sqrt{3}}\qty(\ket{0000}+\ket{0101}+\ket{1010})_{A_1A_2B_1B_2}\\
&=\frac{1}{\sqrt{3}}\qty(\ket{00}+\ket{11}+\ket{22})_{AB}.
\end{align*}
with the identification/relabeling $\ket{0}_A = \ket{00}_{A_1A_2}$, $\ket{1}_A = \ket{01}_{A_1A_2}$, $\ket{2}_A = \ket{10}_{A_1A_2}$, and $\ket{3}_A=\ket{11}_{A_1A_2}$ (similarly for Bob's systems).
We find this to be an additional indication of the similarity existing between the theory of coherence and the theory of entanglement, and of the role that the (generalized) $\ent{CNOT}$ plays in the mapping between coherence and entanglement~\cite{streltsov_measuring_2015,theurer_resource_2017,regula_converting_2017} as well as between general quantumness (of correlations) and entanglement~\cite{piani_all_2011,gharibian_characterizing_2011, piani_quantumness_2012}

\section{Conclusions}

We have introduced a quantifier of genuine distributed coherence for multipartite systems. It is based on the combination of a quantifier of distributed coherence---the gap between global and local coherences---together with a minimization of such a quantifier over all possible global incoherent unitaries. This is justified by the fact that in principle, in the framework established by \cite{baumgratz_quantifying_2014}, and considering the natural choice of global incoherent basis as product of the local incoherent bases, incoherent unitaries that permute, up to a phase, elements of such a global basis are `free'. We remark that there is an on-going debate about the right class of `incoherent operations' that should be considered as free, in particular taking into account that one can distinguish between speakable and unspeakable notions of coherence~\cite{marvian_how_2016,chitambar_critical_2016,streltsov_colloquium_2017}. The class of unitaries we consider as free makes the theory developed in this paper be about speakable coherence. Nonetheless, the starting quantifier $\dc$ of distributed coherence is well-defined also in other frameworks, and one could define alternative measures of genuine distributed coherence minimizing over other meaningful classes of unitaries. If one such class was to be either a subset or a superset of the class of incoherent unitaries we do consider in the present paper, then our genuine-distributed-coherence quantifier would play the role of upper bound or lower bound, respectively, on such said alternative quantifier of genuine distributed coherence. Given that the class of unitaries \eqref{Eq:IncoherentUnitary} is the most general that preserves incoherent states, our quantifier is more likely to play the role of lower bound in future studies in other resource-theoretic frameworks. For that matter, we like to imagine that our approach to quantify genuine distributed coherence may contribute to the discussion about the validity and consistency of alternative resource-theoretic frameworks.

Even staying within our framework, plenty of problems and possible venues of research stay open. While we have given numerical evidence that the state \eqref{eq:maximalGDC} is the two-qubit state that exhibits the largest value of genuine distributed coherence $\gdc$, an analytical proof is lacking. Also, obviously, it would be good to generalize our detailed analysis also to higher dimensions, mixed states, and multipartite systems. In particular, with respect to the point of considering multipartite systems, we observe that our quantifiers $\dc$ and $\gdc$ do not distinguish between (genuine) distributed coherence that involves only a limited number of parties, and coherence that involves many or all parties. This ties with another notion of `genuine' multipartite properties, like `genuine multipartite entanglement'~\cite{acin_classification_2001,coffman_distributed_2000,dur_three_2000,guhne_entanglement_2009,horodecki_quantum_2009}, where one cares whether all parties contribute to the property simultaneously. While this kind of concept was explored in, e.g.,  \cite{radhakrishnan_distribution_2016}, we believe more could be done, even just at the level of distributed coherence like the one captured by $\dc$. One could then still further consider the issue of redistributing coherence by means of incoherent global unitary transformations. Going in this direction, one could study a sort of `supergenuine multipartite coherence', where one takes into account at the same time both the issue of how many parties contribute to the coherence, and of the ability to redistribute coherence `freely' by means of incoherent unitaries.

Finally, there is the open question of what kind of effects / uses may be related to genuine distributed coherence, as well as of the means to detect such form of coherence, e.g., by means of suitably defined witnesses, like it can be done for coherence and multilevel coherence~\cite{napoli_robustness_2016,piani_robustness_2016,ringbauer_quasi-device-independent_2017}.

\section*{Acknowledgements}

We acknowledge support by the Foundational Questions Institute under the Physics of the Observer Programme (Grant No. FQXi-RFP-1601). We thank Xiao-Dong Yu for useful discussions, and Gerardo Adesso for feedback on a preliminary version of this manuscript.
Part of this work was conducted during the 657$^\mathrm{o}$ WE-Heraeus Seminar “Quantum Correlations in Space and Time”, and we gratefully acknowledge the support and the hospitality of the WE-Heraeus Foundation and of the Physikzentrum Bad Honnef.

\end{document}